\documentclass[aps,twocolumn,prb,tightenlines,showpacs]{revtex4}

\usepackage{graphicx}
\usepackage[english]{babel}
\usepackage{amsmath}
\usepackage{amssymb}
\usepackage{times}
\usepackage {appendix}

\usepackage {epstopdf}

\newcommand{\be}{\begin{eqnarray}}
\newcommand{\ee}{\end{eqnarray}}
\newcommand{\p}{\partial}

\newcommand {\rmd} {{\rm d}}
\newcommand{\dbar}{{\rm d}\mkern-6mu\mathchar'26}  

\newcommand{\uxx}{\partial_xu^x}
\newcommand{\uxy}{\partial_xu^y}
\newcommand{\uyx}{\partial_yu^x}
\newcommand{\uyy}{\partial_yu^y}
\newcommand{\vu}{\mathbf{u}}

\begin{document}

\title{Negative Thermal Expansion and Some Elastic properties of a Class of Solids}

\author{Yan He, Vladimir Cvetkovic, C.\ M.\ Varma}

\affiliation{Department of Physics, University of California, Riverside, CA}

\date{\today}

\begin{abstract}
We consider the thermal expansion, change of sound velocity with pressure and
temperature, and the Poisson ratio of lattices which have rigid units (polyhedra
very large stiffness to change in bond-length and to bond-angle variations) connected
to other such units through relatively compressible polyhedra. We show that in such a model, 
the potential energy for rotations of the rigid units can occur only as a function of the combination  
 ${\boldsymbol \Theta}_i \equiv \left ( {\boldsymbol \theta}_i - (\nabla \times {\bf u}_i) /2 \right )$, where 
${\boldsymbol \theta_i} $ are the orthogonal rotation angles of the rigid unit $i$
and ${\bf u}_i$ is its displacement.  Given such new invariants in the theory of elasticity and the hierarchy of force constants of the model,
a negative thermal expansion coefficient as well as a decrease in the elastic constants of the solid with temperature and
pressure is shown to follow. These are consistent with the observations.
\end{abstract}


\maketitle


\section{Introduction}

Most solids expand on raising temperature. There has been a large interest recently in
materials that do the opposite. A particular example is ZrW$_2$O$_8$ in which the
volume decreases linearly with temperature from about 10 K to about 1000 K. The 
thermal expansion coefficient $\alpha = (1/V) \partial V/\partial T \approx - 2.6 \times 10^{-5} K^{-1}$ and
is isotropic over this range \cite{Drymiotis, MarySleight, EvansSleight}.
Other examples are discussed by Tao and
Sleight \cite{TaoSleight}. These solids appear to share the feature that they contain
`rigid units'. By this is meant that the complicated lattice structure of such solids is
composed of corner or edge sharing polyhedra which are very stiff towards internal
bond-stretching or bond-bending compared to the stiffness for change of the relative
angles of the polyhedra and to other elastic forces which determine the sound-velocities.

It was proposed very early \cite{MarySleight, EvansSleight} that the  dynamics of rigid units must be of primary importance in negative thermal
expansion in such solids. One of the first models
proposed is one in which each polyhedra is connected to others by three
or more polyhedra. In this case, rotation of any polyhedra must be accompanied
by the rotation of the whole solid. Rotation of individual polyhedra, which was
the idea behind the negative thermal expansion is then not a thermodynamic
variable. Perhaps, to circumvent this problem, a model was introduced \cite{heine} in
which the atoms connecting the polyhedra was split in to two with a spring in
between them. The analysis of such a model was carried out by standard methods of evaluating the dynamical
matrix with some assumed potential functions. This gave what were termed "rigid unit modes" throughout the
 Brillouin zone which appear not to agree with the spectra measured by neutron
scattering \cite{neutrons} or with the temperature dependence of the specific
heat in ZrW$_2$O$_8$ \cite{Schlesinger2} which appear anomalous only at low temperatures.

\begin{figure}
\includegraphics[width=0.4\textwidth]{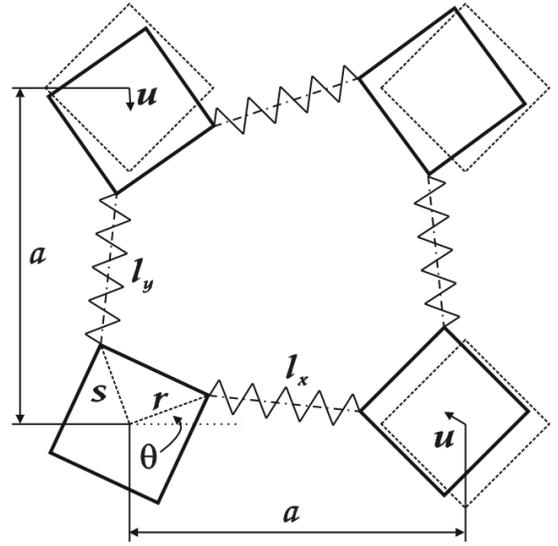}
\caption{The unit cell in our problem. In the ground state at $T = 0$ the distance between the
neighboring squares is $a$ (lattice constant). The displacement of the center of each square $i$ is
${\bf u}_i$, and its rotation from the equilibrium is expressed through angle $\theta_i$. The
rods ${\bf l}_\alpha$ connect neighboring squares and they are very rigid. The other rigidity
is due to the varying angle between rods and squares.}
\label{Fig_square}
\end{figure}

A variant of the model \cite{SimonVarma}, represented in two dimensions by Fig.\ 1 of Ref.\
[\onlinecite{SimonVarma}] for
such crystals was proposed. The essential feature of this model is
that the polyhedra are of two kinds, those that are three or more-fold connected and
those that are only two fold connected. In such a model the two fold connected squares can be replaced by rigid rods.  The generalization of this to three dimensions
is also given in Ref.\ [\onlinecite{SimonVarma}] and has the same properties as the
two-dimensional model. The polyhedra in ZrW$_2$O$_8$, for example satisfy such properties. The two dimensional model is the same as that represented by Fig.\ \ref{Fig_square} here, with the springs between squares replaced by fixed rods. 

In Fig.1 of Ref.\ [\onlinecite{SimonVarma}] the squares do not change their shape of size and the only degree
of freedom in the unit-cell is the angle $\theta$. The area (volume in 3d) occupied
by the solid is the sum of the area (volume) occupied by the solid squares (polyhedra)
and the area (volume) between them. The latter are bounded by fixed perimeters
(areas). Under this condition, the area (volume) is maximum when the perimeter
(surface) is most symmetric. Any distortion from the condition of equilibrium 
leads to change in $\theta$ with a change in area (volume) $\propto (\delta \theta)^2$. Since
the thermal average $ \langle (\delta \theta)^2 \rangle \propto T$ for $T \gtrsim \omega_0$, the
bond-bending energy, a decrease in volume with temperature proportional to
temperature is to be expected. 

The model of Fig.\ 1 of Ref.\ [\onlinecite{SimonVarma}], though adequate to describe the physics of thermal
expansion, is however not a good representation of a solid. The
only degree of freedom is the rotation angle $\theta$. The non-linearity of the problem
leads to an interesting mathematical problem but there there are no elastic modes
in such models which all solids must have. A simple modification of the model
removes this serious shortcoming and in fact makes the problem solvable much more easily.

The modification is to replace the rigid sticks or the rigid two-fold connected
polyhedra with sticks or polyhedra with finite compressive stiffness to obtain the
model represented in Fig.\ \ref{Fig_square}. It is important
that this stiffness be large compared to the energy to change $\theta$. As we will
show, this hierarchy of stiffnesses --- very large (so as to be taken to be infinity
or rigidly constrained) for all internal co-ordinates of the four-fold connected
squares compared to that of the spring which is in turn is large compared to
the stiffness for changing $\theta$ preserves the feature which gives
negative $\alpha$  in the earlier model. It has the virtue of having finite elastic
constants required of a solid so that sound velocity as a function of volume or pressure
or temperature can also be calculated and compared with experiments. Recent experiments show that  solids with $\alpha < 0$ also have a decrease in elastic constants both with temperature and with pressure \cite{Drymiotis, PanteaMigliori}. We will also be able to explain this behavior. We will also show that these properties are consistent with the general relation between the Gr\" uneisen parameter $\gamma$ which gives the change in vibration frequencies with volume and $\alpha$.

The two dimensional model is exhibited in Fig.\ \ref{Fig_square}. Just as in the
previous paper \cite{SimonVarma}, a three dimensional model with the same properties can be
shown to have similar properties. Instead of just one rotational degree, they have three rotational degrees of freedom per unit-cell. But the primary feature, the new invariant in the energy, exhibited in Eq.\
\eqref{invariant} and its coupling to other degrees of freedom, has similar properties in the there-dimensional model. Our intent in this paper is only to exhibit the special features introduced by the hierarchy of the stiffnesses in solids with rigid units and not to calculate the elastic properties quantitiatively.

We should add that the physical features of the model are similar to those in the model of previous investigators \cite{TaoSleight, MarySleight, EvansSleight, heine}. Our formulation of the problem reveals new general physical and mathematical features of the problem. Some numerical work has also been done on a model similar to that treated here \cite{Schlesinger2}. 

\section{Equations of Motion and Normal Modes}

Let us count the degrees of freedom per unit-cell in the model. There are 8 {\it total} degrees
of freedom per unit-cell and five rigid constraints which fix the lengths and the internal
angles of the squares. That leaves 3 dynamical degrees of freedom per unit-cell.
(In the earlier model \cite {SimonVarma}, the length of the rods was also fixed
introducing two more constraints thus leaving only one degree of freedom, $\theta$ per
unit-cell.) The three degrees of freedom may be taken to be $\{{\bf u}_{i}, \theta_{i}\}$
where ${\bf u}_{i}= ({\bf R}_{i} - {\bf R}_{i}^{(0)}) \equiv (u_{i}^{x} \hat{x} + u_{i}^{y} \hat{y})$
is the displacement of the center of the mass of the square labelled $(ij)$ from its
equilibrium position at $T = 0$. The two extra degrees of freedom compared to
the earlier model ensure elastic properties of the model, the $\theta_i$ degree
of freedom provides an optic mode which is crucial as before for the anomalous
thermal properties for $T$ comparable or larger than its energy.

Let us focus on a unit cell. The geometric position of the squares and springs
are shown in the schematic figure Fig.\ \ref{Fig_square}. For convenience, we
consider a single square at position $i$, and find its free energy. The summation over the entire
lattice is then assumed.

The two neighboring corners of squares $i$ and $i + \hat {x}$ are spaced apart by vector
\be
\label{constraint}
  {\bf l}_{i, \hat x} = a {\hat x} + {\bf u}_{i + \hat x} - {\bf u}_i - a \tfrac \xi 2 (\hat r_i + \hat r_{i + \hat x}). \label {l1x}
\ee
Unit vectors $\hat r_i = (\cos \theta_i, \sin \theta_i)$ and $\hat s_i = (- \sin \theta_i, \cos \theta_i)$
point along diagonals of square $i$, and $\xi$ is the ratio of square diagonal to the square (i.e., lattice)
spacing $a$. The vector ${\bf l}_{i, \hat x}$ corresponds to the rod between squares $i$ and $i + \hat x$.
Analogously, for the other three corners/rods we have
\be
  {\bf l}_{i, \hat y \phantom {-}} &=& \phantom {-} a {\hat y} + {\bf u}_{i + \hat y} - {\bf u}_i - a \tfrac \xi 2 (\hat s_i + \hat s_{i + \hat y}), \nonumber \\
  {\bf l}_{i, - \hat x} &=& -a {\hat x} + {\bf u}_{i - \hat x} - {\bf u}_i + a \tfrac \xi 2 (\hat r_i + \hat r_{i - \hat x}), \label {other_l} \\
  {\bf l}_{i, - \hat y} &=& -a {\hat y} + {\bf u}_{i - \hat y} - {\bf u}_i + a \tfrac \xi 2 (\hat s_i + \hat s_{i - \hat y}). \nonumber
\ee
The nominal length of each of the rods at $T = 0$ is $l^{(0)} = a (1 - \xi)$.

In this model there are two contributions to the potential energy. There is a rotational
stiffness due to bending between the rods and a square; this is parametrized
by an angle $\mu_{i, \alpha}$ defined through ($\alpha = \pm \hat x$ or $\pm \hat y$)
\be
  \cos \mu_{i, \pm \hat x} &=& \pm \frac {\hat r_{i} \cdot {\bf l}_{i, \pm \hat x}}{|| {\bf l}_{i, \pm \hat x} ||}, \nonumber \\
  \cos \mu_{i, \pm \hat y} &=& \pm \frac {\hat s_{i} \cdot {\bf l}_{i, \pm \hat y}}{|| {\bf l}_{i, \pm \hat y} ||}. \label {cos_mu}
\ee
The other stiffness is due to the tendency of the rods to preserve their nominal length $l^{(0)}$ and
it is proportional to the elastic constant of the rods $g$ that is taken to be large compared to the rotational stiffness. 
The total potential energy is
\be
\label{poten}
  V = \sum_{i, \alpha} \left \lbrack K (1 - \cos \mu_{i, \alpha}) +
    \frac g 2 (|| {\bf l}_{i, \alpha} || - l^{(0)})^2 \right \rbrack. \label {Vpot}
\ee

The potential energy Eq.\ \eqref  {Vpot} needs to be augmented with the kinetic term. Assuming
each square has mass $M$, the kinetic energy reads
\be
  T = \frac M 2 \sum_i \left \lbrack {\dot {\bf u}_i}^2 + a^2 \xi^2 J (\dot \theta_i)^2 \right \rbrack. \label {Tkin}
\ee
The second term, the rotational energy, is proportional to the moment of inertia $M (a \xi)^2 J$.
The constant $J$ is a dimensionless number depending on the mass distribution in each
square. For example, the uniform mass distribution yields $J = 1/3$, while mass concentrated at
the square corners results in $J = 1$.
 
In Appendix A, we calculate the potential energy for long wave-length vibrations up to the fourth order in the
displacements ${\bf u}$ and rotations $\theta$.. 
The harmonic part of the potential energy, Eq.(\ref{poten}), at  long wavelengths is
\be
V_{h}=a^2g\Big[\kappa\Theta^2+\frac{1}{2}(u_{xx}^2+u_{yy}^2)+\kappa u_{xy}^2\Big] \label {harmonic}
\ee
where $\kappa = 2 K / (a^2 g (1 - \xi)^2)$ is a dimensionless parameter.
Also,
\be
\label{invariant}
\Theta = \Big(\theta-\frac{1}{2}\nabla\times\vu\Big).
\ee
The second and the third terms in Eq.\ \eqref{harmonic} are the familiar terms from the theory of elasticity, where we
also introduced strain fields  \cite{LandauL}
\be
  u_{ij} \equiv \frac 1 2 \left ( \frac {\partial u^i}{\partial x_j} + \frac {\partial u^j}{\partial x_i} +
    \frac {\partial u^k}{\partial x_i} \frac {\partial u^k}{\partial x_j} \right ). \label {uij}
\ee
The first term in Eq.\ \eqref{harmonic} is unique to the model. A term in energy proportional
to $({\bf \nabla} \times {\bf u})^2$ is
forbidden because a global rotation of the solid should cost no energy. However, a term proportional to $\Theta^2$, as derived in the appendix, is obviously allowed. In fact the interaction expanded to any order in the displacement can depend on $\theta$ and $\nabla \times {\bf u}$ only in the combinations ${\Theta}$, as explicitly shown in the appendix to cubic and quartic order. This is the basic new feature of the vibrations of solids with rigid units.  All the interesting new results of this
paper can be traced to this feature.

Arranging the degrees of freedom into a column $\psi = (\theta, u^x, u^y)$ the equations of
motion to harmonic order can be concisely written as
\be
M {\rm diag}[Ja^2\xi^2,1,1]~ \p_t^2\psi +M_V\psi= 0. \label {eqmotion_harmonic}
\ee
In the long wavelength limit, we have
\be
M_V=\frac{a^2g}{2}\left(\begin{array}{ccc}
2\kappa & \kappa iq_y & -\kappa iq_x \\
-\kappa iq_y & -q_x^2-\kappa q_y^2 & 0\\
\kappa iq_x & 0 & -\kappa q_x^2-q_y^2
\end{array}\right)
\ee
It is easy to verify that
\be
M_V\left(\begin{array}{c}
1\\-iq_y/2\\iq_x/2
\end{array}\right)\approx a^2g\kappa\left(\begin{array}{c}
1\\-iq_y/2\\iq_x/2
\end{array}\right)
\ee
Then the above vector is an eigenvector and we can rewrite 
\be
\left(\begin{array}{c}
1\\-iq_y/2\\iq_x/2
\end{array}\right)^{\dagger}\psi=\theta+iq_yu^x/2-iq_xu^y/2=\Theta
\ee
The corresponding eigenvalue is $\omega^2=
2a^2g\kappa/(Ma^2\xi^2J)=\frac{2g\kappa}{JM\xi^2}$.
There are also two gapless elastic modes (linear combinations of $u_{x,y}$) required of a two dimensional lattice
 with eigenvalues $\sim \frac{a^2g}{2M}q_{x,y}$.

Setting $q_x = q\,\cos\phi$ and $q_y = q\,\sin\phi$,
the dispersion up to $q^2$ can be written as
\be
  (\omega_{1,2})^2 &=& \frac {a^2 g}{4 M} \left \lbrack 2 + \kappa \pm \right . \label {omega12} \\
  && \left . \sqrt {(2 - 2 \kappa + \kappa^2) + 2 (1 - \kappa) \cos 4 \phi } \right \rbrack q^2, \nonumber  \\
  (\omega_{3 \phantom {4}})^2 &=& \frac {2 g \kappa}{J M \xi^2} \left \lbrack 1 +
    \tfrac {\xi (-2 + \xi + 2 J \xi)}{8} a^2 q^2 \right \rbrack =  \label {omega3}  \\
  && \frac {4 K}{J M a^2 \xi^2 (1 - \xi)^2}
    \left \lbrack  1 +  \tfrac {\xi (-2 + \xi + 2 J \xi)}{8} a^2 q^2 \right \rbrack. \nonumber
\ee
In order to obtain the $q^2$ term in Eq.\ \eqref {omega3}
we need to use the subleading long-wavelength expansion of the potential 
energy given in Appendix B.
Up to this order, the dispersion of the optical mode, which is the mode of the $\Theta$ degree of freedom
Eq.\ \eqref {omega3} is isotropic and does not depend on the spring constant $g$.

The elastic modes have a velocity anisotropy consistent with the underlying lattice
symmetry, and correspond to a square lattice with leading order elastic constants
\be
  C_{11}^0 = g, \quad C_{12}^0 = 0, \quad C_{44}^0 = \frac {g \kappa} 2 = \frac {K}{a^2 (1-\xi)^2}. \label {Cs}
\ee
It should be noted that  we have constructed a very simple model to highlight the new features due to the dynamics of rigid units and kept only the bare minimum of rigidities
relevant for the anomalous thermal properties.  Because we chose only nearest neighbor central force constants $C_{12}^0 =0.$
We can therefore make comparison to
experiments only for the elastic constants $C_{11}$ and $C_{44}$.

We will consider corrections to these elastic constants at finite temperature and with changing pressure in Sec. IV.

The Poisson ratio of the model follows  from the effective elastic constants Eq.\ \eqref {Cs}.
For $C_{12}^0 =0$ the Poisson ratio is \cite{poisson}
\be
  \nu_{2D} &=& \frac {\left ( C_{11}^0 - 2 C_{44}^0 \right ) \sin^2 2 \phi }
    { C_{11}^0 (1 + \cos^2 2 \phi) + 2 C_{44}^0 \sin^2 2 \phi} \nonumber \\
  &=& \frac {( 1 - \kappa ) \sin^2 2 \phi}{ (1 + \cos^2 2 \phi) + \kappa \sin^2 2 \phi} . \label {Poisson}
\ee
When $\kappa < 1$, as it is in our case, the Poisson ratio is strictly positive. Conversely,
when $\kappa > 1$ it is always negative. So, it appears that not all solids with constraints have auxetic behavior
(i.e., $\nu < 0$) \cite {Rechtsman, Bruinsma}.

We can compare this spectrum of this model with the rigid constraint model.
If we assume the angle $\theta$ is small and the difference of displacement is high order
than angles $({\bf u}_{i+\alpha}-{\bf u}_{i})=O(\theta_i^2)$ for $\alpha=\pm\hat{x},\,\pm\hat{y}$,
we can approximately solve the constraint $||{\bf l}_{i,\alpha}||=a(1-\xi)$. We find
\be
  u^x_{i\pm\hat{x}}-u^x_i &=& \frac{- a\xi}{8(1-\xi)}\Big[(2-\xi)(\theta_i^2+\theta_{i\pm\hat{x}}^2)
  +2\xi\theta_i\theta_{i\pm\hat{x}}\Big]\nonumber\\
   \label {solve_ux} \\
  u^y_{i\pm\hat{y}}-u^y_i &=& \frac{-a\xi}{8(1-\xi)}\Big[(2-\xi)(\theta_i^2+\theta_{i\pm\hat{y}}^2)
  +2\xi\theta_i\theta_{i\pm\hat{y}}\Big] . \nonumber\\
  \label {solve_uy}
\ee
These solutions correspond to the solution Set II from Ref.\ [\onlinecite {SimonVarma}].
Plugging these into the expression for the angles, we have
\be
  1-\cos\mu_{i,\alpha}&\approx&\frac{1}{4(1-\xi)^2}\Big([2+(-2+\xi)\xi](\theta_i^2+\theta_{i+\alpha}^2)\nonumber\\
  & &+2(-2+\xi)\xi\theta_i\theta_{i+\alpha}\Big) . \label {cos_mu_exact}
\ee
Then we find the equation of motion for $\theta$ is
\be
  M a^2 \xi^2 J \omega^2 \theta_i = \frac{4 K}{(1-\xi)^2} \left \lbrack \theta_i
  + \frac {(\xi - 2) \xi} 8 \sum_{\alpha}  (\theta_i-\theta_{i+\alpha})\right \rbrack . \label {eqmotion_exact}
\ee
The resulting dispersion is
\be
  \omega^2=\frac {K}{J M} \frac {4}{a^2\xi^2(1-\xi)^2}\left\lbrack1+\tfrac {\xi(-2+\xi)}{8} (a{\bf q})^2\right\rbrack
  \label {omega_exact}
\ee
We compare this result to the optical mode Eq.\ \eqref {omega3}. The gaps
are the same, however, the long wavelength dispersion differs slightly. We notice, however,
that if we were to take first the limit $g \to \infty$, and $\kappa \to 0$ of the harmonic terms
 in a controlled way such that their product $K = \kappa g a^2 (1 - \xi)^2 /2$
remains finite, the dispersions coincide.

\section {Change in Volume with Temperature}

So far we have  only considered the harmonic approximation for the new model. At this level, the
thermal expansion must be zero. The change of volume
$\delta V$ is proportional to the divergence of the
displacement field. In harmonic approximation:
\be
  \delta V/V = \langle \nabla \cdot {\bf u} \rangle \approx
    \langle u_{ii} \rangle = 0. \label {deltaS}
\ee
Here and after the summation over indices is assumed in term $u_{ii} = u_{xx} + u_{yy}$.

In order to get non-zero $\delta V$, we must consider anharmonic 
terms in our Hamiltonian which we find from the expansion of Eq.\ \eqref {Vpot} given in the Appendix.
The cubic anharmonic term is 
\be
\label{anharmonic}
  V_{\rm pot}^{\rm (3)} &=& \int \rmd {\bf r} ~ \frac {g}{ 2 (1 - \xi) } \Big \lbrace \xi
    (u_{ii}) \Theta^2 - 2 \xi \Theta (u_{xx} - u_{yy}) (u_{xy}) + \nonumber \\
  && \quad \quad \xi (u_{xy})^2 (u_{ii}) - (1 - \xi) \left \lbrack (u_{xx})^3 + (u_{yy})^3
    \right \rbrack \Big \rbrace.  \label {Vanh}
\ee

If we were to keep all the anharmonic terms the thermal expansion would depend on the
thermal averages $\langle \Theta^2 \rangle_{T}$, $\langle (u_{ij}) \Theta \rangle_T$,
and $\langle (u_{ij}) (u_{kl}) \rangle_T$. We know, however, that in
the physical limit that we are interested in ($\kappa \ll 1$), the first average
is of order $O (\kappa)^{-1}$, while the others are of order $O (\kappa)^0$.
Thus they are smaller by a factor of $\kappa$, which allows us to
concentrate on the first term in Eq.\ \eqref {Vanh}.
For the same reason, terms proportional to $\kappa$ were omitted in Eq.\ \eqref {Vanh}
and will be omitted hereafter in all other calculations.
Adding this anharmonic term to the Lagrangian the equations of motion
(at $\omega = 0$) become
\be
  g \partial_\alpha^2 u^\alpha = - \frac {g \xi}{2 (1 - \xi)} \partial_\alpha (\Theta^2). \label {eqmotion_anh}
\ee
Index $\alpha$ my be either $x$ or $y$.  Eq.\ \eqref {eqmotion_anh} is equivalent to the
Eqs.\ (\ref {solve_ux} - \ref {solve_uy}) derived earlier from consideration of the constraints alone.

Integrating Eq.\ \eqref {eqmotion_anh} and taking its average we arrive to
\be
  \langle \nabla \cdot {\bf u} \rangle (T)  =
    - \frac {\xi}{1 - \xi} \langle \Theta_{\bf x}^2 \rangle (T) . \label {divu_theta2}
\ee
The thermal average of $\Theta^2$ is given below, and the classical limit is
\be
  \langle \Theta_{\bf x}^2 \rangle (T) &=& \int \dbar {\bf q} \frac 1 \beta \sum_{\omega_n}
    \frac {1}{M a^2 \xi^2 J \left \lbrack (i \omega_n)^2 - (\omega_3 ({\bf q}))^2 \right \rbrack} \nonumber \\
  &=& \int \dbar {\bf q} \frac {1}{2 M a^2 \xi^2 J \omega_3 ({\bf q})}
    \coth \frac {\beta \omega_3 ({\bf q})}{2} \nonumber \\
  &\stackrel {k_B T \gg \omega_3}{\longrightarrow} & \int \dbar {\bf q}
    \frac {k_B T}{M a^2 \xi^2 J (\omega_3 ({\bf q}))^2} \nonumber \\
  & \approx& \frac {k_B T}{2 a^2 g \kappa} + O (\kappa)^0 . \label {theta2avg}
\ee
In the last step we assumed that the massive optical mode is only weakly dispersive. The
thermal expansion coefficient obtained by combining Eqs.\ \eqref {divu_theta2} and
\eqref {theta2avg} is
\be
  \alpha = - \frac {k_B \xi}{2 a^2 g \kappa (1 - \xi)} = - \frac {k_B \xi (1 - \xi)}{4 K}. \label {thermexp}
\ee

The physics of the negative thermal expansion coefficient is the same
as in the earlier model and discussed in the Introduction. The usual anharmonicity of the longitudinal and the
transverse acoustic modes
cannot change this because in the hierarchy of the assumed stiffnesses, such
modes have low occupation compared to the rotation modes of the rigid units because their energy is higher than
 $\omega_3$, Eq.\ \eqref {omega3} over most of the Brillouin zone. The thermal contraction (as well as the change in sound velocity with temperature derived below) comes from the first anharmonic term in Eq. (\ref{anharmonic}) which in turn can be traced to the fundamental constraint, Eq. (\ref{constraint}) which links the rotation freedom of the rigid units which costs low energy to the motion of the degrees of freedom ${\bf u}_{i}$ which determine the volume of the lattice.

\section{Change in Velocity of Sound with Temperature, Pressure or Volume}

The change in the elastic constants with temperature at fixed pressure of $ZrW_2O_8$
\cite{Drymiotis} is similar to those of the normal solids which expand on increasing
temperature, i.e., they decrease on increasing temperature. At a fixed temperature the elastic constants also decrease with pressure.
It is worth discussing in general that the two properties are mutually consistent and consistent with the relation of the Gr\" uneisen parameter $\gamma$ and the thermal expansion coefficient.

The relation
\be
\label{thermo}
(\frac{\partial \omega}{\partial V})_T = \frac{(\partial \omega/\partial P)_T}{(\partial V/ \partial P)_T }
\ee
gives that since $({\partial V/ \partial P})_T <0$ for stability, decrease in the vibration frequencies $\omega$ with pressure must be accompanied by an increase in $\omega$ with volume. $\gamma_{\omega} \equiv (- \frac{\partial \omega}{\partial V})_T $ is therefore negative. The thermal expansion coefficient is shown in general to be proportional to $\gamma$ defined to be an average over all $\omega$ of $\gamma_{\omega}$ , on thermodynamic grounds \cite{ashcroft-mermin}. It thus follows that the the observed behavior of the change in volume with temperature, and the change in elastic constants with pressure are mutually consistent.  However, we proceed to show this explicitly by calculating both the change in the elastic constants with temperature and with pressure. 

Consider the cubic anharmonic  terms in Eq.\ \eqref {Vanh}.  As seen above,
a change in temperature leads to a nonzero expectation value for strains at finite
 temperature
 \be
   \langle u_{ij} \rangle (T) = \tfrac 1 2 \alpha T \delta_{ij}. \label {avg_delu}
 \ee
Therefore term such as $u_{xx}$, and $u_{yy}$ in Eq.\ \eqref {Vanh} can be replaced
by their expectation value. This leads to corrections to the harmonic part
of the potential \cite {Kleinert2}.  This correction is linear in $T$:
\be
  V^{(3)} (T) &\to& \int \rmd {\bf r} ~ \frac {g}{ 2 (1 - \xi) } \Big \lbrace \xi \Theta^2 +
     \xi (u_{xy})^2 - \label {VpT} \\
  && \quad \quad 3 \frac {1 - \xi}{2} \left \lbrack (u_{xx})^2 + (u_{yy})^2 \right \rbrack
    \Big \rbrace   \langle \nabla \cdot {\bf u} \rangle (T) . \nonumber
\ee
Small correction due to $\kappa$ are ignored in Eq.\ \eqref {VpT}. The first term of Eq.\ \eqref {VpT} yields small corrections
(proportional to $T$) to the mass of the optical mode and can therefore be neglected in
our work. Only the second and the third term contribute to renormalization of the
elastic constants Eq.\ \eqref {Cs}.

In addition to Eq.\ \eqref {VpT}, there is another term of the same order in
temperature affecting the dispersion of elastic modes.  It is derrived from quartic anharmonic term and
the fact that the average $\langle \theta^2 \rangle_T$ also depends on temperature. Expanding potential
Eq.\ \eqref {Vpot} up to the forth order and keeping only $O(\kappa)^0$ terms, we find
\be
  V_{\rm pot}^{\rm (4)} &=& \int \rmd {\bf r} ~ \frac {a g \xi}{4 (1 - \xi)^2} \Big \lbrace \xi \Theta^4
     + \Theta^2 \Big \lbrack  6 \xi (u_{xy})^2  \nonumber \\
  && \quad \quad - (1+\xi) \lbrack (u_{xx})^2 + (u_{yy})^2 \rbrack \Big \rbrack + \ldots \Big \rbrace . \label {Vanh4}
\ee
Here we have introduced a parameter $a$ to fix the relative magnitude of the quartic to the cubic anharmonicity. In the simple model that we have introduced, this parameter is necessary to get the right variation of all the elastic constants with temperature as well as pressure.
Terms of kind $\Theta (\partial u)^3$ and $(\partial u)^4$ lead to higher order corrections and
have been dropped. Substituting the thermal average $\langle \Theta^2_{\bf x} \rangle_T$
in Eq.\ \eqref {Vanh4} we obtain another correction to the harmonic potential
\be
  V^{(4)} (T) &\to& \int \rmd {\bf r} ~ \frac {a g \xi \langle \Theta^2 \rangle_T}{4 (1 - \xi)^2}
    \Big \lbrace 2 \xi \Theta^2 + 6 \xi (u_{xy})^2 - \nonumber \\
  &&  \quad \quad - (1 + \xi) \lbrack (u_{xx})^2 + (u_{yy})^2 \rbrack  \Big \rbrace . \label {VppT}
\ee
Same as in Eq.\ \eqref {VpT}, the first term only affects the optical mode mass and is therefore
neglected.

Combining the two corrections Eqs.\ \eqref {VpT} and \eqref {VppT} to potential,
the temperature dependence of effective elastic constants  is found
\be
  \left ( \frac {\partial C_{11}}{\partial T} \right )_P &=&
    -g\frac{(3 - a -\xi(3 +a))}{2(1-\xi)} \alpha, \label {C11T} \\
  \left ( \frac {\partial C_{44}}{\partial T} \right )_P &=&g \frac {(1-3a) \xi}{1 - \xi} \alpha, \label {C44T}
\ee
where $\alpha$ is the (negative) thermal expansion coefficient found in Eq.\ \eqref {thermexp}.

The pressure dependence of the elastic constants at a constant temperature is
derived in an analogous fashion. 
\be
  \left ( \frac {\partial C_{11}}{\partial P} \right )_T &=& g\frac{(3-a) -(3+a)\xi}{2(1-\xi)} \kappa_T, \label {C11p} \\
  \left ( \frac {\partial C_{44}}{\partial P} \right )_T &=& -g \frac {(1-3a) \xi}{1 - \xi} \kappa_T, \label {C44p}
\ee
where $\kappa_T$ is the isothermal compressibility measuring change
in volume proportional to applied pressure: $(\delta V) / V = - \kappa_T (\delta P)$.
Within the present model 
\be
  \langle \nabla \cdot {\bf u} \rangle (P)= - \frac {2}{C_{11}^0 + C_{12}^0} P = - \frac 2 g P. \label {div u}
\ee
The change of the elastic constants with pressure and temperature are consistent with the thermodynamic requirement discussed in Eq.\ \eqref {thermo}.

\section{Comparison with normal solids with ${\bf \alpha > 0}$}

It is worthwhile discussing how the differences from solids with normal thermal expansion coefficient $\alpha >0$
come about.
We may consider the following (schematic) long wavelength Hamiltonian for them.
\be
\label{normal H}
H = KE + (1/2) {\bf \partial u \kappa \partial u} + {\bf g}_3 {\bf \partial u \partial u \partial u}.
\ee
 KE is the kinetic energy, ${\bf \kappa}$ is the matrix of harmonic force constants and ${\bf g}_3$ the cubic-anharmonic  tensor.
 An expectation value $\langle {\bf \partial u \partial u} \rangle = w^2(T)$ is finite at all temperatures so that we may decouple the last term and minimize to get that there is a change in volume which may be schematically represented by
 \be
   \langle \partial u \rangle = - g_3 w^2(T)/\kappa. \label {partialu_g3}
 \ee
 Solids usually have $g_3 < 0$.  This ensures that a increase(decrease) of interatomic distance decreases (increases) interatomic interaction energy relative to the harmonic case. This then leads to a positive thermal expansion. This is to be contrasted with the $g$ derived above in the model with rigid units  from the expansion of potential energy subject to the constraint Eq.\ \eqref{constraint}.  In this case {$g_3 > 0$, as also required by the general argument given in the Introduction.

 Given a finite $\langle \partial u \rangle$, we may decouple the second term in Eq.\ \eqref{normal H} also in the alternative way to find an effective stiffness renormalization:
 \be
   \kappa_{eff} = \kappa_0  + 2 g_3 \langle \partial u \rangle = \kappa - 2 g_3^2 w^2(T) / \kappa. \label {kappa_eff}
 \ee
We see that the correction to the effective stiffness does not depend on the sign of $g_3$; we always have a decrease in the elastic constants with temperature due to the cubic anharmonic term since the mean square displacement $w^2(T)$ must always increase with temperature consistent with entropy always increasing with temperature. However the quartic term acts in an opposite direction in general and to get elastic softening their effect should be smaller than that of the cubic anharmonicity. As we see below to get this feature of the experiments we have to introduce the parameter $a \ll 1$.

\section {Comparison with the experiments}

Zirconium tungstate is about the most studied materials with negative expansion
coefficient. We now check whether our calculations of the thermal expansion coeffcient Eq.\ \eqref {thermexp},
and the temperature (Eqs.\ \eqref {C11T} and \eqref {C44T}) and pressure (Eqs.\ \eqref {C11p}
and \eqref {C44p}) corrections to the elastic constants 
are in qualitative accord with the experiments.

In our model a negative thermal expansion coefficient $\alpha$, linear in temperature occurs for temperatures above the energy of the rotation optic mode. In experiments, $\alpha \propto - T$ above about $10$ K.  We can estimate the parameter $K \approx  100 {\rm Kelvin}$ using an effective mass of $O(500)$ atomic units, to
get $\omega_3 \approx 10 K$. 

The thermal expansion coefficient $\alpha = -2.6 \times 10^{-5} {\rm K}^{-1}$.  Our result yields this order of magnitude with $K \approx 10^2 K$ for $\xi \approx 0.9$. 

The relative decrease of the elastic constant $(1/C_{11}^0)\partial C_{11}(T)/\partial T$ is $O(10^{-3})$
and of $(1/C_{44}^0)\partial C_{44}(T)/\partial T$ is an order of magnitude smaller
\cite{Drymiotis}. Similarly the decrease of $C_{44}$ with pressure is an order of
magnitude smaller than that of $C_{11}$ \cite{PanteaMigliori}. To get the
ratio between these quantities, we need to use $a \approx 1/4$ in Eqs.\ (\ref {C44T}) and (\ref {C11p}). 

With these values, we calculate $(1/C_{11}^0)\partial C_{11}(T)/\partial T $ of the right sign but about $5$ times larger in magnitude.  This indicates the perils of trying to get the experimental numbers through a model which is too simple in terms of details despite our introduction of the phenomenological parameter $a$. Part of the difficulty stems from our ignoring normal anharmonic terms to concentrate on the new physics due to the rigid unit rotations. The usual anharmonic terms are always used as adjustable parameters anyway.

\section {Conclusions}

We have constructed a model of a solid with negative thermal expansion coefficient.
The solid consists of rigid units interconnected by elastic units where the elastic stiffness
is much higher than the next relevant stiffness related to the bond bending. The
model has sensible elastic properties even though the optical mode is identical to that
found earlier in fully constrained models, which did not have that virtue. A novel feature is the uncovering of a new invariant
in the theory of elasticity in this class of models. This may be of relevance also in other contexts. 

We do obtain the correct qualitative features for the thermal expansion coefficient and the ratio of the change in the temperature and pressure dependence of the two elastic constants we have calculated. The quantitative agreement cannot be aspired for, although We obtain the thermal expansion coefficient, which depends on the details less than the elastic constants, quite well quantitatively with reasonable parameters. We have had to introduce an ad-hoc parameter $a$, the ratio of the cubic to the quartic anharmonic parameter, to estimate the change of elastic constants with temperature and pressure quantitatively. Despite that this simple model gives the quantitative value only to within a factor of about 5 although the relative change in the two elastic constants calculated is given much better.

\begin{widetext}

\appendix

\section {Expansion of the Potential Energy}

In this appendix section, we show the details of the calculation of the derivation of  the expansion of the potential energy into harmonic and leading anharmonic terms.
Since we assume the stiffness of the spring is much larger than the rotation, we will focus on the expansion of
\be
  V =\sum_{i,\alpha}\frac g 2(|| {\bf l}_{i, \alpha} || - l^{(0)})^2
  \label{vg}
\ee
In order to show the result in a rotational invariant form we introduce $\Theta$ and strains:
\be
\Theta=\theta-\frac{1}{2}\nabla\times\vu,\quad\quad u_{ij}=\frac{1}{2}\Big[\p_iu^j+\p_ju^i+(\p_iu^k)(\p_ju^k)\Big]
\ee

The nonlinear term in the strains is important for proving that at each expansion order the potential energy is rotational invariant. It is straightforward to expand Eq. (\ref{vg}) to find the quadratic terms $V_2=\frac{a^2g}{2}\Big[(\uxx)^2+(\uyy)^2\Big]$. The cubic and fourth order potential terms are given as follows.
\be
V_3&=&\frac{a^2g}{2(1-\xi)}\Big[\xi\theta^2(\uxx+\uyy)+2\xi\theta(\uyx\uyy-\uxy\uxx)
+\uxx(\uxy)^2+\uyy(\uyx)^2\Big]\nonumber\\
V_4&=&\frac{a^2g}{4(1-\xi)^2}\bigg(\xi^2\theta^4+2\xi^2\theta^3(-\uxy+\uyx)
+\xi\theta^2\Big[(1+2\xi)[(\uxy)^2+(\uyx)^2]
-2\xi[(\uxx)^2+(\uyy)^2]\Big]+\cdots\bigg)\nonumber
\ee
Here $\cdots$ in $V_4$ are $\Theta(\p u)^3$ and $(\p u)^4$ type of terms which lead to higher order corrections.
Plug $\theta=\Theta+\frac{1}{2}(\uxy-\uyx)$ into $V_3$, then rearrange terms, we find
\be
V_3&=&\frac{a^2g}{2(1-\xi)}\Big[\xi\Theta^2(\uxx+\uyy)
-\xi\Theta(\uxy+\uyx)(\uxx-\uyy)\nonumber\\
& &\quad\quad+\frac{\xi}{4}(\uxy+\uyx)^2(\uxx+\uyy)\Big]
+\frac{a^2g}{2}[\uxx(\uxy)^2+\uyy(\uyx)^2]\label{v31}
\ee
Now we want to write this in terms of strains. But we have to remember there are cubic order contributions from $V_2$.
we find
\be
  V_2=\frac {a^2g}{2} \left \lbrack (u_{xx})^2 + (u_{yy})^2 \right \rbrack - \frac {a^2g}{2}
    \left \lbrack (\partial_x u^x)^3 + (\partial_x u^x) (\partial_x u^y)^2 + (\partial_y u^y)^3 +
    (\partial_y u^x) (\partial_y u^y)^2 \right \rbrack + O^4 (\partial u)
    \label{V2u}
\ee
In Eq.(\ref{V2u}), $O^4 (\partial u)$  denote the 4th or higher order terms of $\p u$. The second term of Eq.(\ref{V2u}) should be combined with $V_3$.
\be
&&V_3=\frac{a^2g}{2(1-\xi)} \left \lbrack \xi\Theta^2u_{ii}
-2\xi\Theta(u_{xx}-u_{yy})u_{xy}+\xi u_{xy}^2u_{ii}\Big]
-\frac{a^2g}{2}[u_{xx}^3+u_{yy}^3 \right \rbrack+O^4(\p u,\Theta)
\ee
Here $O^4(\p u,\Theta)$ are 4th or higher order terms in $\p u$ and $\Theta$ which we ignored in $V_3$ and $u_{ii}=u_{xx}+u_{yy}$.

We can go on to rewrite $V_4$ in terms of $\Theta$ and strains. We should include the 4th order contribution from  $V_2$ and $V_3$. But for $V_4$, we only need terms proportional to $\Theta^4$, $\Theta^3$ and $\Theta^2$, thus the only correction terms we need is the following
$$
-\frac{a^2g\xi}{4(1-\xi)}\Theta^2\Big[(\uxx)^2+(\uxy)^2+(\uyx)^2+(\uyy)^2\Big]
$$
which is coming from $\xi\Theta(\uxx+\uyy)$ term in Eq. (\ref{v31}). Then after some algebraic calculation, we find
\be
V_4=\frac{a^2g\xi}{4(1-\xi)^2}\bigg(\xi\Theta^4
+\Theta^2\Big[6\xi u_{xy}^2-(1+\xi)(u_{xx}^2+u_{yy}^2)\Big]+\cdots\bigg)+O^5(\p u,\Theta)
\ee
These contributions are rewritten in Sec. III. However, to compare with experiments, it is found necessary in 
Sec.\ III to introduce a phenomenological parameter $a$, which is the ratio of the quartic to the cubic anharmonic potentials.

\section {Subleading long wavelength expansion of the potential energy}

In this section we present the gradient expansion of potential energy Eq.\ \eqref {Vpot}
where the next order terms in lattice spacing $a$ are not truncated. These terms are
relevant for finding the weak dispersion of the optical mode in Eq.\ \eqref {omega3}
and the comparison of our model to the fully constrained model at the end of Sec.\ II.
For all other purposes, these terms can be truncated and the harmonic potential
in Eq.\ \eqref {harmonic} are sufficient.

Using earlier defined phase space vector $\psi$, the potential energy is written as
\be
  V_{pot }= \frac {a^2 g} 2 ~ \left ( \begin {array}{c} \theta \\ u^x \\ u^y \end {array} \right )
   \left ( \begin {array} {c c c}
    2 \kappa \left \lbrack 1 - \frac {\xi (2 - \xi) a^2 \partial_i^2} 8 \right \rbrack &
    \kappa \left \lbrack \partial_y - \frac {a^2 \partial_y^3}6 \right \rbrack &
    -\kappa \left \lbrack \partial_x - \frac {a^2 \partial_x^3}6 \right \rbrack \\
    - \kappa \left \lbrack \partial_y - \frac {a^2 \partial_y^3}6 \right \rbrack &
    - \left \lbrack \partial_x^2 + \frac {a^2 \partial_x^4}{12} \right \rbrack -
    \kappa \left \lbrack \partial_y^2  + \frac {a^2 \partial_y^4}6 \right \rbrack & 0 \\
    \kappa \left \lbrack \partial_x - \frac{a^2 \partial_x^3}6 \right \rbrack & 0 &
    - \left \lbrack \partial_y^2 + \frac {a^2 \partial_y^4}{12} \right \rbrack  -
    \kappa \left \lbrack \partial_x^2 + \frac {a^2 \partial_x^4}6 \right \rbrack 
  \end {array} \right )
  \left ( \begin {array}{c} \theta \\ u^x \\ u^y \end {array} \right ) \label {Vpotsub}
\ee

In the constrained model, the limiting
procedure where $\kappa \to 0$ and $g \to \infty$, in such way that their product (and
therefore $K$) remains finite yields the dispersion of Eq.\ \eqref {omega_exact}.

\end{widetext}

\bibliographystyle{apsrev}

\end{document}